\documentclass[aps,prd,reprint,nofootinbib,floatfix]{revtex4-2}
\IfFileExists{newtxtext.sty}{
  \usepackage[T1]{fontenc}
  \usepackage{newtxtext}
  \usepackage{newtxmath}
}{
  \usepackage{mathptmx}
}

\usepackage{amsmath,amssymb,bm}
\usepackage{graphicx}
\usepackage{booktabs}
\usepackage{xcolor}
\usepackage[colorlinks=true,linkcolor=blue,citecolor=magenta,urlcolor=blue]{hyperref}
\AtBeginDocument{\let\hbar\hslash}

\def\doi#1{\href{https://doi.org/#1}{\color{blue}#1}}
\usepackage{tikz}
\usetikzlibrary{shapes.geometric}
\DeclareRobustCommand{\orcidicon}{
  \begin{tikzpicture}
    \definecolor{orcidgreen}{HTML}{166B3A}
    \draw[orcidgreen,fill=orcidgreen] (0,0)
    circle [radius=0.16]
    node[white] {{\fontfamily{qag}\selectfont \tiny ID}};
  \end{tikzpicture}
  \hspace{-2mm}
}
\foreach \x in {A, ..., Z}{
  \expandafter\xdef\csname orcid\x\endcsname{\noexpand\href{https://orcid.org/\csname orcidauthor\x\endcsname}{\noexpand\orcidicon}}
}

\newcommand{\pr}{p_r}
\newcommand{\pt}{p_t}
\newcommand{\Mpl}{M_{\mathrm{Pl}}}
\newcommand{\Rnn}{R_{99}}
\newcommand{\dd}{\mathrm{d}}
\begin{document}
\raggedbottom

\title{Intrinsic pressure anisotropy in spherical Proca stars}

\author{Il{\'i}dio Lopes\,\orcidA{}}
\affiliation{\href{https://ror.org/020hp3377}{Centro de Astrof{\'i}sica e
Gravita\c{c}\~ao (CENTRA)}, Departamento de F{\'i}sica, \\
Instituto Superior T\'ecnico (IST),
\href{https://ror.org/01c27hj86}{Universidade de Lisboa (UL)}, \\
Av.~Rovisco Pais~1, 1049-001 Lisboa, Portugal}

\date{August 2026}

\begin{abstract}
Pressure anisotropy in relativistic stars is commonly prescribed through a
phenomenological closure, obscuring its microscopic origin and relation to
stress-energy conservation. Here it is derived directly from the minimally
coupled Einstein--complex-Proca theory. For spherical Proca stars, the
principal-pressure difference admits an exact on-shell form whose sign is
controlled solely by the local mass-shell threshold. The stress is radially
dominated in the core, reverses on a surface of fixed gravitational redshift,
and becomes tangentially dominated in the envelope. At the first mass maximum,
the fractional anisotropy reaches about \(21\%\) near the density maximum,
whilst the region beyond the reversal contains about \(9\%\) of the mass. Its
exact atmospheric limit is approximately \(24\%\), equal in magnitude and
opposite in sign to the scalar-boson-star limit. Because the usual local fluid
variables remain
non-zero at the crossing, no sign-definite closure constructed from them can
reproduce the profile. These results identify an intrinsically vectorial stress
reversal, generated without additional interactions, and provide a
first-principles benchmark for anisotropic bosonic compact objects.
\end{abstract}

\maketitle

\section{Introduction}
\label{sec:introduction}

A massive vector boson, light enough to behave as a classical field on
astrophysical scales, is amongst the more economical candidates for cold dark
matter. Such a field is produced in the observed abundance either by vacuum
misalignment with a non-minimal coupling to curvature
\citep{NelsonScholtz2011}, or by inflationary fluctuations, whose abundance is
fixed by the vector mass and the inflationary scale alone
\citep{GrahamEtAl2016}; it need not couple to the Standard Model at all. On
galactic scales a field so light behaves as a wave rather than as a gas, and
relaxes into solitonic cores at the centres of haloes \citep{Hui2021};
simulations of the vector case find such cores too, and they carry a
polarisation structure that the scalar case does not possess
\citep{AminEtAl2022}. The relativistic, self-gravitating realisation of one of
those cores is a Proca star, and it is the stress structure of the minimal
such star that concerns me here.

Fully nonlinear Einstein--complex-Proca stars, both spherical and rotating,
were first constructed a decade ago by Brito et al. \citep{BritoEtAl2016}. In
the spherical
family, the geometry is static although the complex vector potential is
harmonic in time, whilst the global \(U(1)\) symmetry yields a conserved
Noether charge. That phase symmetry requires a complex field: a single real
Proca field has no corresponding charge, and self-interacting real vectors
form intrinsically time-dependent oscillons instead \citep{ZhangJainAmin2022}.
Nonlinear evolutions later settled the stability of the spherical branch
\citep{SanchisGualEtAl2017}, and the construction has since been carried to
rotating, hairy, self-interacting, multipolar and binary settings
\citep{HerdeiroEtAl2016,HerdeiroRadu2020,
LazarteAlcubierre2024,SanchisGualEtAl2019}. Black-hole superradiance constrains
part of the ultralight-vector mass range \citep{BaryakhtarEtAl2017}, a merger
of two Proca stars has been advanced as one reading of a gravitational-wave
transient \citep{CalderonBustilloEtAl2021}, and the lensing, light rings and nonlinear
dynamics of the wider bosonic-star family have been surveyed in detail
\citep{CunhaEtAl2017,LieblingPalenzuela2023,HerdeiroEtAl2024}.

Unequal principal pressures, \(\pr\ne\pt\), are in stellar interiors closer to
the rule than to the exception, and they arise in several distinct ways. At
supranuclear density nuclear matter may itself become anisotropic
\citep{Ruderman1972}, a condensed pion phase \citep{Sawyer1972} being the most
often quoted realisation. A magnetised fermion gas separates its longitudinal
and transverse pressures by the field-magnetisation term
\citep{FerrerEtAl2010}. A mixture of two fluids that do not move together is
exactly equivalent to a single anisotropic fluid \citep{Letelier1980}, a
mechanism realised in the dark sector whenever a bosonic component is carried
alongside the baryonic one \citep{BurasStubbsLopes2024}. The isotropic
condition is not merely one possibility amongst many but is itself unstable,
dissipation, density inhomogeneity and shear driving an initially isotropic
configuration away from isotropy \citep{Herrera2020}. Nor are the consequences
small: the extra force density \(2(\pt-\pr)/r\) displaces mass--radius curves
and compactness bounds appreciably
\citep{BowersLiang1974,HerreraSantos1997,RaposoEtAl2019}.

What is almost always absent is a derivation. In practice the anisotropy is
closed by hand, and the closure is chosen for tractability: the Bowers--Liang
form \citep{BowersLiang1974}, a quasi-local dependence on density and
compactness \citep{HorvatEtAl2011}, the vanishing of Herrera's complexity
factor, applied to exotic matter in Ref.~\citep{RinconEtAl2023}, a covariant
construction fitted to an ultracompact regime \citep{RaposoEtAl2019}, or, for
stars made of dark matter, an interior solution with the anisotropy assigned
at the outset \citep{MoraesEtAl2021,PanotopoulosEtAl2025}. Each is a
defensible modelling choice. None, however, identifies the degrees of freedom
that carry the stress, nor follows from the dynamics of the matter it purports
to describe.

Bosonic stars are free of that arbitrariness, for nothing in them is closed by
hand. Their stresses follow from a covariant action, their profiles are solved
together with the metric, and the field equations imply \(\nabla_\mu
T^{\mu\nu}=0\) for the matter stress tensor \(T^{\mu\nu}\) without further
assumption. Scalar boson stars have a long history
\citep{Kaup1968,RuffiniBonazzola1969,LieblingPalenzuela2023}, and their
anisotropy, recognised from the earliest studies \citep{Gleiser1988}, has been
standard ever since \citep{SchunckMielke2003,AlcubierreEtAl2022}. For the
vector, separated stress components have appeared in a \(3+1\) formulation
\citep{SanchisGualEtAl2017,LazarteAlcubierre2024}, but their difference has
never been reduced to closed form.

We report three apparently new results and one direct consequence. First, the
on-shell pressure anisotropy reduces to a closed expression whose sign is
carried by the local mass-shell function. It is negative wherever a static
observer measures the field frequency above the vector mass. Second, it
reverses on the fixed-redshift surface where the locally measured frequency
equals that mass, a single radius on the fundamental branch. Third, the
fractional anisotropy approaches an exact frequency-dependent limit in the
atmosphere, equal in magnitude and opposite in sign to the scalar result.
Because radius, density, radial pressure, and enclosed compactness remain
positive at the crossing, these results also exclude any sign-definite closure
built only from those variables: the Proca anisotropy is governed by the lapse,
rather than by local fluid data alone.

The remainder of the paper is organised as follows.
Section~\ref{sec:model} introduces the covariant Einstein--Proca model and its
spherical reduction. Section~\ref{sec:anisotropy} derives the field-generated
pressure anisotropy and its sign criterion. Sections~\ref{sec:numerics}
and~\ref{sec:results} describe the numerical construction and present the stellar
sequence, characteristic radii, stress profiles and radial stability boundary,
respectively. Section~\ref{sec:discussion} discusses the implications for
phenomenological closures and the scalar counterpart, whilst
Sec.~\ref{sec:conclusions} summarises the main results.
Here and below, radial stability refers only to stability against spherical
perturbations.
I adopt standard general-relativistic conventions
\citep{MisnerThorneWheeler1973}: the metric signature is \((-,+,+,+)\), and
natural units are used with \(\hbar=c=1\). Newton's constant is denoted by
\(G\), and \(\Mpl\equiv G^{-1/2}\) is the unreduced Planck mass. Greek indices
run over spacetime; \(\mu\) standing alone, without an index position, denotes
the vector mass.

\section{Covariant Einstein--Proca model}
\label{sec:model}

\subsection{Action, current, and field equations}
\label{sec:action}

The massive-vector theory originates with Proca \citep{Proca1936}, and early
gravitational analyses of the minimally coupled Einstein--Proca system may be
found in Refs.~\citep{Bekenstein1972,ObukhovVlachynsky1999}. For a complex
Proca one-form \(\mathcal A_\mu\), and with \(\mathfrak
g\equiv\det(g_{\mu\nu})\), the minimal action is
\citep{BritoEtAl2016,SanchisGualEtAl2017}
\begin{equation}
S=\int \dd^4x\,\sqrt{-\mathfrak g}\left[
\frac{\mathcal R}{16\pi G}
-\frac14\mathcal F_{\lambda\tau}\overline{\mathcal F}^{\lambda\tau}
-\frac12\mu^2\mathcal A_\lambda\overline{\mathcal A}^{\lambda}
\right],
\label{eq:action}
\end{equation}
where \(\mathcal R\) is the Ricci scalar, \(\mathcal
F_{\mu\nu}=\nabla_\mu\mathcal A_\nu-\nabla_\nu\mathcal A_\mu\) is the field
strength, \(\mu\) is the vector-field mass, and an overbar denotes complex
conjugation. Variation with respect to the metric and to \(\overline{\mathcal
A}_\mu\) gives, with \(G_{\mu\nu}\) the Einstein tensor,
\begin{align}
G_{\mu\nu}&=8\pi G T_{\mu\nu},
\label{eq:einstein}\\
\nabla_\mu\mathcal F^{\mu\nu}&=\mu^2\mathcal A^\nu .
\label{eq:proca}
\end{align}
Since \(\nabla_\nu\nabla_\mu\mathcal F^{\mu\nu}\) vanishes identically, the
divergence of Eq.~\eqref{eq:proca} yields the Lorenz relation
\begin{equation}
\nabla_\mu\mathcal A^\mu=0,
\label{eq:lorenz}
\end{equation}
here a dynamical consequence of \(\mu\ne0\) rather than a gauge choice. The
mass term removes the local gauge redundancy; the complex field then carries
three propagating polarisations, that is, six real degrees of freedom, and
retains a global phase symmetry. With the matter Lagrangian density written
\(\mathcal L_m\), the Hilbert stress tensor \citep{BritoEtAl2016}, \(T_{\mu\nu}=-2(-\mathfrak
g)^{-1/2} \delta[(-\mathfrak g)^{1/2}\mathcal L_m]/\delta g^{\mu\nu}\), is
\begin{align}
T_{\mu\nu}={}&
\frac12\left(
\mathcal F_{\mu\lambda}\overline{\mathcal F}_{\nu}{}^\lambda+
\overline{\mathcal F}_{\mu\lambda}\mathcal F_{\nu}{}^\lambda
\right)
-\frac14g_{\mu\nu}
\mathcal F_{\lambda\tau}\overline{\mathcal F}^{\lambda\tau}
\nonumber\\
&+\frac{\mu^2}{2}\left(
\mathcal A_\mu\overline{\mathcal A}_\nu+
\overline{\mathcal A}_\mu\mathcal A_\nu
-g_{\mu\nu}\mathcal A_\lambda\overline{\mathcal A}^{\lambda}
\right),
\label{eq:stress}
\end{align}
and diffeomorphism invariance of the matter action guarantees \(\nabla_\mu
T^{\mu\nu}=0\) on any solution of Eq.~\eqref{eq:proca} and its conjugate,
independently of Eq.~\eqref{eq:einstein}; the contracted Bianchi identity then
renders the two statements consistent.

The global \(U(1)\) phase symmetry yields the conserved current
\citep{BritoEtAl2016,HerdeiroRadu2020}
\begin{equation}
j^\mu=\frac{i}{2}\left(
\overline{\mathcal F}^{\mu\nu}\mathcal A_\nu-
\mathcal F^{\mu\nu}\overline{\mathcal A}_\nu
\right),
\qquad \nabla_\mu j^\mu=0,
\label{eq:current}
\end{equation}
where conservation follows from Eq.~\eqref{eq:proca}. The corresponding
dimensionless Noether charge is
\(Q_{\rm phys}=-\int_\Sigma j^\mu n_\mu\sqrt{h}\,\dd^3x\), where \(n^\mu\) is
the future-directed unit normal to the spacelike hypersurface \(\Sigma\), and
\(h\) is its induced-metric determinant. After quantisation, \(Q_{\rm phys}\)
counts net \(U(1)\) quanta, or bosons on the positive-charge branch.

\subsection{Spherical reduction}
\label{sec:reduction}

The static, spherically symmetric line element, in the mass-function form of
Refs.~\citep{Tolman1939,OppenheimerVolkoff1939}, and the compatible Proca
potential, whose harmonic phase generalises the complex-scalar ansatz of
Refs.~\citep{Kaup1968,RuffiniBonazzola1969}, are
\begin{align}
\dd s^2={}&-\sigma^2(r)N(r)\dd t^2+\frac{\dd r^2}{N(r)}
+r^2\dd\Omega^2,\nonumber\\
N(r)={}&1-\frac{2m(r)}{r},
\label{eq:metric}\\
\mathcal A={}&e^{-i\omega t}\left[f(r)\dd t+i g(r)\dd r\right],
\label{eq:ansatz}
\end{align}
where \(m(r)\) is the real mass function, \(\sigma\), \(f\) and \(g\) likewise
real, \(\omega\) is the bound-state frequency, and \(\dd\Omega^2\) is the
unit-sphere metric. Every bilinear entering Eq.~\eqref{eq:stress} is then time
independent, the configuration being invariant under a combined time
translation and phase rotation. The factor~\(i\) is essential: it annihilates
the mixed mass bilinear \(\mathcal A_t\overline{\mathcal
A}_r+\overline{\mathcal A}_t\mathcal A_r\) and, with it, the radial charge
flux. The remaining mixed bilinear \(\mathcal F_{t\lambda}\overline{\mathcal
F}_r{}^\lambda\) vanishes for any purely electric configuration, whatever the
phase of \(g\); the energy flux \(T_{tr}\) therefore vanishes, the stress
tensor is diagonal in the static frame, and \(\pr\), \(\pt\) are genuine
principal stresses. The reduced system used here is that of the
pioneering Proca-star construction \citep{BritoEtAl2016}; its Proca equations
are
\begin{align}
\frac{\dd}{\dd r}\left[
\frac{r^2(f'-\omega g)}{\sigma}\right]
&=\frac{\mu^2r^2f}{\sigma N},
\label{eq:proca1}\\
\omega g-f'&=\frac{\mu^2\sigma^2Ng}{\omega},
\label{eq:proca2}
\end{align}
where a prime denotes \(\dd/\dd r\). Two of the Einstein equations are
\begin{align}
m'={}&4\pi G r^2\left[
\frac{(f'-\omega g)^2}{2\sigma^2}
+\frac{\mu^2}{2}\left(
Ng^2+\frac{f^2}{N\sigma^2}\right)\right],
\label{eq:mass}\\
\frac{\sigma'}{\sigma}={}&
4\pi G r\mu^2\left(
g^2+\frac{f^2}{N^2\sigma^2}\right).
\label{eq:sigma}
\end{align}
The square bracket in Eq.~\eqref{eq:mass} is the local energy density
\(\rho\equiv T_{\mu\nu}u^\mu u^\nu=-T^t{}_t\), measured by the static observer
with unit four-velocity
\(u^\mu=(\sigma\sqrt N)^{-1}\delta^\mu{}_t\). Thus Eq.~\eqref{eq:mass}
takes the familiar mass-balance form \(m'=4\pi G r^2\rho\).

The Lorenz condition supplies the remaining first-order equation,
\begin{equation}
g'=-\frac{\omega f}{\sigma^2N^2}
-g\left(\frac{2}{r}+\frac{\sigma'}{\sigma}+\frac{N'}{N}\right),
\label{eq:gprime}
\end{equation}
and Eqs.~\eqref{eq:mass}--\eqref{eq:gprime}, together with the algebraic
relation of Eq.~\eqref{eq:proca2}, form the radial system. With \(f_c=f(0)\)
and \(\sigma_c=\sigma(0)\) in an arbitrary central-time gauge, regularity at
the origin gives \citep{BritoEtAl2016}
\begin{align}
f(r)&=f_c+\frac{f_c}{6}
\left(\mu^2-\frac{\omega^2}{\sigma_c^2}\right)r^2
+\mathcal O(r^4),\nonumber\\
g(r)&=-\frac{f_c\omega}{3\sigma_c^2}r+\mathcal O(r^3),
\nonumber\\
m(r)&=\frac{4\pi G f_c^2\mu^2}{6\sigma_c^2}r^3
+\mathcal O(r^5),\nonumber\\
\sigma(r)&=\sigma_c+
\frac{4\pi G f_c^2\mu^2}{2\sigma_c}r^2+\mathcal O(r^4).
\label{eq:centre}
\end{align}
At large radius, \(m(r)\to m_\infty\), and the asymptotic-time gauge is fixed
by \(\sigma\to1\), so that \(\omega\) is the frequency measured at infinity
and \(N(r)=1-2m_\infty/r\) up to exponentially small terms. Linearising the
Proca equations about this Schwarzschild tail
\citep{GaltsovEtAl1984,Konoplya2006} gives, up to relative corrections of
order \(r^{-1}\),
\begin{equation}
\begin{gathered}
f(r)\sim C_f e^{-\kappa r}r^p,
\qquad g(r)\sim\frac{\omega}{\kappa}f(r),\\
\kappa=\sqrt{\mu^2-\omega^2},
\qquad 0<\omega<\mu,\\
p=-1+\frac{m_\infty(2\omega^2-\mu^2)}{\kappa}.
\end{gathered}
\label{eq:bound}
\end{equation}
Exponential localisation selects the decaying branch; at \(\omega=\mu\) the
exponential scale disappears, whereas \(\omega>\mu\) produces an oscillatory,
non-localised field. Gravity therefore leaves the decay rate \(\kappa\)
unchanged but shifts the algebraic index from its flat-space value \(p=-1\).
The atmospheric e-folding length is \(\kappa^{-1}\), so configurations become
increasingly diffuse as \(\omega\to\mu^-\).

Comparison with the asymptotic Schwarzschild metric gives
\(M_{\rm ADM}=m_\infty/G\). Since a solitonic configuration has no sharp
material surface, its effective radius \(R_x\) is defined by
\(m(R_x)=x m_\infty/100\); as \(r\) is an areal coordinate, the corresponding
sphere has area \(4\pi R_x^2\). In what follows,
\(\Rnn\equiv R_x|_{x=99}\) denotes the areal radius enclosing 99 per cent of
the ADM mass.

\section{Field-derived anisotropy}
\label{sec:anisotropy}

In the static observer's orthonormal frame, and with lapse
\(\alpha=\sigma\sqrt N\), I define
\begin{equation}
\mathcal E^2=\frac{(f'-\omega g)^2}{\sigma^2},\qquad
\mathcal X_0=\frac{\mu^2f^2}{\alpha^2},\qquad
\mathcal X_r=\mu^2Ng^2.
\label{eq:pieces}
\end{equation}
The energy density just defined and the principal pressures follow from
Eq.~\eqref{eq:stress},
\begin{align}
\rho&=\frac12\left(\mathcal E^2+\mathcal X_0+\mathcal X_r\right),
\label{eq:rho}\\
\pr&=\frac12\left(-\mathcal E^2+\mathcal X_0+\mathcal X_r\right),
\label{eq:pr}\\
\pt&=\frac12\left(\mathcal E^2+\mathcal X_0-\mathcal X_r\right),
\label{eq:pt}
\end{align}
whence, on eliminating \(f'-\omega g\) by means of Eq.~\eqref{eq:proca2},
\begin{equation}
\Delta\equiv\pt-\pr=\mathcal E^2-\mathcal X_r
=\mu^2Ng^2\left(\frac{\mu^2\alpha^2}{\omega^2}-1\right).
\label{eq:delta}
\end{equation}
The individual stresses are known and the elimination is a single
substitution; the on-shell form nevertheless yields a transparent criterion
for a sign that is neither imposed nor independently adjustable. Where
\(g\ne0\) and \(N>0\), that sign rests on the last bracket alone. Since
\(\omega/\alpha\) is the proper frequency measured by a static observer,
\(\Delta<0\) exactly where that frequency exceeds \(\mu\), and the principal
pressures exchange rank on the surface
\begin{equation}
\alpha=\frac{\omega}{\mu},
\label{eq:turning}
\end{equation}
which is the local mass-shell threshold. Equivalently, Eq.~\eqref{eq:delta}
reads
\begin{equation}
\Delta=-\frac{\mu^2\alpha^2Ng^2}{\omega^2}
\left[\left(\frac{\omega}{\alpha}\right)^{\!2}-\mu^2\right].
\label{eq:wavenumber}
\end{equation}
Here the bracket is the local mass-shell function: in a WKB description it is
positive for propagation, zero at threshold, and negative for evanescence.
Curvature and spherical-measure terms modify the exact Liouville-normal-form
potential, but not this algebraic sign criterion for \(\Delta\).

The lapse gradient follows directly from the Einstein equations.
Differentiating \(\alpha=\sigma\sqrt N\), and using \(N=1-2m/r\) together with
Eqs.~\eqref{eq:mass},~\eqref{eq:sigma}, and~\eqref{eq:rho}--\eqref{eq:pr}, gives
\((\ln\alpha)'=(m+4\pi G r^3\pr)/(r^2N)\). The density terms cancel, leaving
the familiar active-mass combination of enclosed mass and radial pressure.

At its first mass maximum, the fundamental solution has nodeless \(g\), a
single node in \(f\), and \(\alpha(0)<\omega/\mu\). Although \(\pr\) becomes
negative in the envelope, the computed lapse remains monotonically
increasing. Regularity gives \(g=\mathcal O(r)\), and hence \(\Delta=0\) at the
centre; away from it, \(\Delta\) changes sign exactly once, at the local
threshold of Eq.~\eqref{eq:turning}. A non-central crossing requires
\(\alpha(0)<\omega/\mu\) and is unique when \(g\) is nodeless. In excited
states, nodes of \(g\) introduce further zeros of \(\Delta\), where the
principal pressures touch without exchanging rank.

In the exponential atmosphere \(f'\to-\kappa f\) and \(\alpha\to1\), so
Eq.~\eqref{eq:proca2} gives \(g\to\omega f/\kappa\), whence
\(\rho\to\mu^4f^2/\kappa^2\) whilst \(\pr/\rho\to0\) and
\(\pt/\rho\to\kappa^2/\mu^2\). The principal spatial stress thus becomes
asymptotically tangential, even as all absolute stresses decay, and
\begin{equation}
\lim_{r\to\infty}\frac{\Delta}{\rho}
=\lim_{r\to\infty}\frac{\pt}{\rho}=\frac{\kappa^2}{\mu^2}
=1-\frac{\omega^2}{\mu^2}.
\label{eq:tail}
\end{equation}

With \(N>0\) throughout, the energy conditions may be read off directly:
\begin{align}
\rho-\pr&=\mathcal E^2,&
\rho+\pr&=\mathcal X_0+\mathcal X_r,\nonumber\\
\rho-\pt&=\mathcal X_r,&
\rho+\pt&=\mathcal E^2+\mathcal X_0,
\label{eq:dec}
\end{align}
all four being manifestly non-negative, whence also \(\rho\ge0\). Diagonality
having been established above, these inequalities constitute the dominant
energy condition in full, and the strong condition follows, since the
right-hand column gives \(\rho+p_i\ge0\) for \(i=r,t\) and
\(\rho+\pr+2\pt=\mathcal E^2+2\mathcal X_0\). At the origin \(\mathcal E^2\)
and \(\mathcal X_r\) vanish, so that \(\pr=\pt=\rho\) and the dominant energy
condition is saturated at that single point. Radial conservation reduces to
the anisotropic hydrostatic equation
\begin{equation}
\pr'+(\rho+\pr)\frac{\dd}{\dd r}\ln\alpha
-\frac{2\Delta}{r}=0,
\label{eq:atov}
\end{equation}
where \(2\Delta/r\) is the anisotropic force density
\citep{BowersLiang1974,HerreraSantos1997}, and a scalar counterpart of
Eq.~\eqref{eq:atov} is long established \citep{SchunckMielke2003}. With these
identifications the two Einstein equations of Sec.~\ref{sec:reduction} take
their familiar fluid form, the bracket in Eq.~\eqref{eq:mass} being precisely
\(\rho\), whilst Eq.~\eqref{eq:sigma} reads
\(\sigma'/\sigma=4\pi Gr(\rho+\pr)/N\); the geometry is thus sourced by exactly
the stresses that Eqs.~\eqref{eq:rho}--\eqref{eq:pt} extract from the Hilbert
tensor. For the ansatz of Eq.~\eqref{eq:ansatz}, the conserved charge is
\citep{BritoEtAl2016,SanchisGualEtAl2017}
\begin{equation}
Q_{\rm phys}=\frac{4\pi\mu^2}{\omega}
\int_0^\infty r^2\sigma N g^2\,\dd r.
\label{eq:charge}
\end{equation}

\begin{figure*}
	\includegraphics[width=\textwidth]{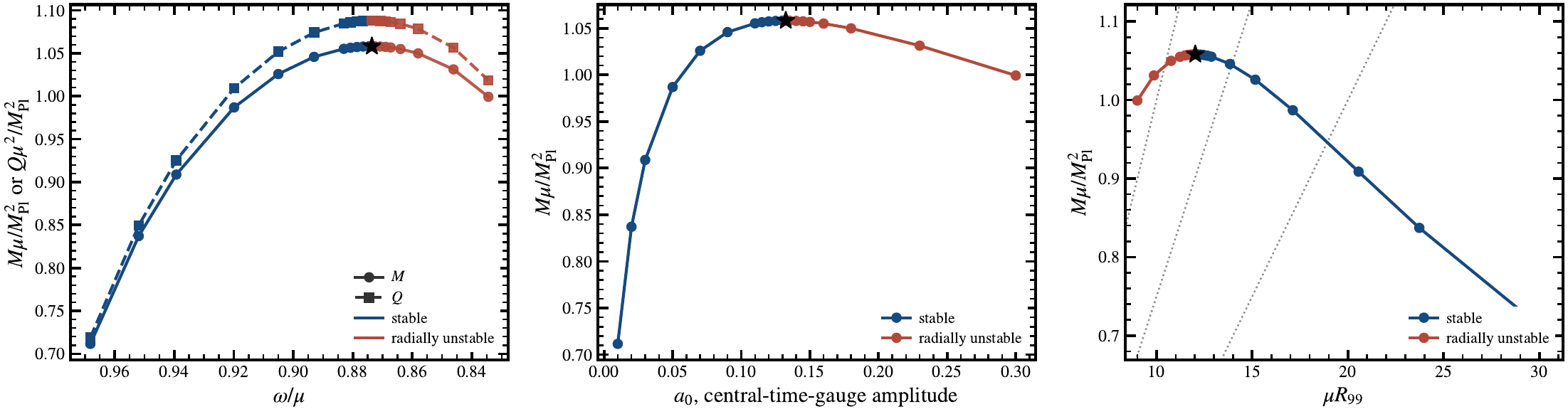}
	\caption{Fundamental spherical Proca-star sequence computed here. Left: ADM
		mass and Noether charge against physical frequency. Centre: mass against
		\(a_0\). Right: the mass--radius relation, with dotted lines of constant
		\(M/\Rnn=0.050\), \(0.075\) and \(0.100\). The star marks the first turning
		point, where \(\dd M/\dd a_0=0\). Locally across it, \(a_0<a_{0,\star}\) is
		radially stable, whilst the adjacent branch \(a_0>a_{0,\star}\) carries one
		unstable spherical mode.}
	\label{fig:sequence}
\end{figure*}

\section{Numerical construction and verification}
\label{sec:numerics}

Dimensionless variables are introduced as \(\widehat r=\mu r\), \(\widehat
m=\mu m\), \(\widehat\omega=\omega/\mu\), \(\widehat f=\sqrt{4\pi G}\,f\) and
\(\widehat g=\sqrt{4\pi G}\,g\), after which the hats are omitted, so that
\(M\), \(Q\) and \(R\) are dimensionless numerical values, related to the ADM
mass, to the charge of Eq.~\eqref{eq:charge} and to physical lengths by
\(M_{\rm ADM}=M\Mpl^2/\mu\), \(Q_{\rm phys}=Q\Mpl^2/\mu^2\) and \(R_{\rm
phys}=R/\mu\). Equation~\eqref{eq:charge} then reads \(Q=\omega^{-1}\!\int
r^2\sigma Ng^2\,\dd r\).

Values and profiles alike come from a global nonlinear eigenvalue
construction. For each fixed invariant amplitude \(a_0=f(0)/\sigma(0)\),
adaptive fourth-order collocation solves simultaneously for \(\bm
y=(m,\sigma,f,g)\), the physical frequency \(\omega\), and the central value
\(\sigma_c=\sigma(0)\) on \([r_0,r_{\max}]\), with \(r_0=10^{-6}\),
\(r_{\max}=35\) and, since \(N(0)=1\), central lapse \(\alpha(0)=\sigma_c\).
The regular central series, asymptotic-time normalisation, and decaying-tail
condition close the two-point problem. Appendix~\ref{app:collocation} sets out
that formulation, and Appendix~\ref{app:shooting} the independent shooting
reconstruction against which it is checked.

Here \(a_0\) is dimensionless, with physical amplitude
\(a_{0,{\rm phys}}=\Mpl a_0/\sqrt{4\pi}\). Regularity gives
\(\widehat\rho_c\equiv4\pi G\rho(0)/\mu^2=a_0^2/2\), whence
\(\dd M/\dd a_0=a_0\,\dd M/\dd\widehat\rho_c\). Thus, for \(a_0>0\), the
central-density and amplitude extrema coincide; \(\dd M/\dd\omega=0\) is
likewise equivalent wherever \(\omega(a_0)\) is locally monotonic, as it is at
the first maximum. We denote its amplitude by \(a_{0,\star}\), with
\(\dd^2M/\dd a_0^2<0\).

The two numerical constructions agree closely and satisfy independent
consistency tests, whose numerical details are collected in
Appendix~\ref{app:shooting}. One such test uses the virial identity, first
obtained for
the free spherical Proca family in Ref.~\citep{BritoEtAl2016} and later
rederived by effective-action methods
\citep{HerdeiroRadu2020,HerdeiroEtAl2021Virial}, which reads
\begin{multline}
\int_0^\infty \dd r\,r^2\sigma\left[
\mu^2\left(g^2-\frac{f^2(4N-1)}{\sigma^2N^2}\right)\right.\\
\left.-\frac{(\omega g-f')^2}{\sigma^2}\right]=0.
\label{eq:virial}
\end{multline}

\subsection{First mass maximum}
\label{sec:sequence}

The collocation sequence reaches its first mass maximum at \(a_0=0.132\),
\(f(0)=0.0971\), \(\omega=0.874\), \(M=1.06\), and \(Q=1.09\),
with central lapse \(\alpha(0)=0.733\). Here and below, a star denotes
evaluation at this maximum; in particular, \(\omega_\star\) is the physical
field frequency, measured at infinity, of the maximum-mass solution.
Table~\ref{tab:benchmark} compares our collocation and shooting results with
two published determinations. The two methods agree at the displayed
precision, with their unrounded differences given in
Appendix~\ref{app:shooting}. Our unrounded values of \(M_\star\) and \(Q_\star\) agree
with Ref.~\citep{BritoEtAl2016} to about \(10^{-4}\). Its quoted frequency
differs more because the maximum was identified on a comparatively coarse
frequency grid; the later spectral value agrees with \(\omega_\star\) to
\(4.6\times10^{-4}\) in relative terms.

At the same turning point, \(\lvert\dd M/\dd a_0\rvert\) and
\(\lvert\dd Q/\dd a_0\rvert\) are both below \(2\times10^{-6}\). The
turning-point criterion \citep{Sorkin1981,Sorkin1982}, supported by radial
analysis \citep{GleiserWatkins1989} and nonlinear Proca-star evolutions
\citep{BritoEtAl2016,SanchisGualEtAl2017}, identifies this extremum as the
boundary at which the first spherical mode is marginal. Locally, the dilute
side \(a_0<a_{0,\star}\) is radially stable, whereas the adjacent branch
\(a_0>a_{0,\star}\) carries one unstable spherical mode. The regular tangent
to the equilibrium family is a useful numerical diagnostic, but, since
\(\dd\omega/\dd a_0\ne0\), it is not the dynamical neutral mode.

\begin{table}
\caption{First mass maximum from global collocation and independent shooting,
compared with the published shooting \citep{BritoEtAl2016} and spectral
\citep{LazarteAlcubierre2024} results. Our columns agree at the displayed
precision; Appendix~\ref{app:shooting} gives their unrounded differences. An
ellipsis denotes a quantity not reported by the reference.}
\label{tab:benchmark}
\centering
\begingroup
\footnotesize
\setlength{\tabcolsep}{3.5pt}
\begin{tabular}{lcccc}
\toprule
& \multicolumn{2}{c}{This work} & &\\
\cmidrule(lr){2-3}
Quantity & Collocation & Shooting & Ref.~\citep{BritoEtAl2016} &
Ref.~\citep{LazarteAlcubierre2024} \\
\midrule
\(\omega_{\star}/\mu\) & 0.8736 & 0.8736 & 0.875 & 0.874 \\
\(M_{{\rm ADM},\star}\mu/\Mpl^{2}\) & 1.058 & 1.058 & 1.058 & 1.058 \\
\(Q_{{\rm phys},\star}\mu^{2}/\Mpl^{2}\) & 1.088 & 1.088 & 1.088 & \(\cdots\) \\
\(\alpha_\star(0)\) & 0.7331 & 0.7331 & \(\cdots\) & 0.733 \\
\(\mu\Rnn\) & 12.02 & 12.02 & \(\cdots\) & 12.03 \\
\bottomrule
\end{tabular}
\endgroup
\end{table}

Figure~\ref{fig:sequence} presents the fundamental family computed here. At
the maximum, \(R_{50}=5.50\), \(R_{90}=8.68\), \(R_{95}=9.75\), and
\(\Rnn=12.02\), so \(M/\Rnn=0.088\). The spectral value
\(M/\Rnn=0.088\) \citep{LazarteAlcubierre2024} corroborates the diffuse
atmosphere, to which \(\Rnn\) is most sensitive. The local compactness peaks at
\(\max_r 2m(r)/r=0.226\) when \(\mu r=7.53\). A photon sphere would require
\(r\,\dd\ln\alpha/\dd r=1\), but the left-hand side never exceeds \(0.151\);
neither a photon sphere nor its associated ringdown therefore occurs. The
ultracompact configurations discussed in Ref.~\citep{CunhaEtAl2017} lie
elsewhere on the family. On the positive-charge branch,
\(M_{\rm ADM}/Q_{\rm phys}=(M/Q)\mu\) is the energy per quantum, so
\(1-M/Q=0.0275\) is the fractional binding energy against free dispersion,
not proof of dynamical stability.

\begin{figure*}
	\includegraphics[width=\textwidth]{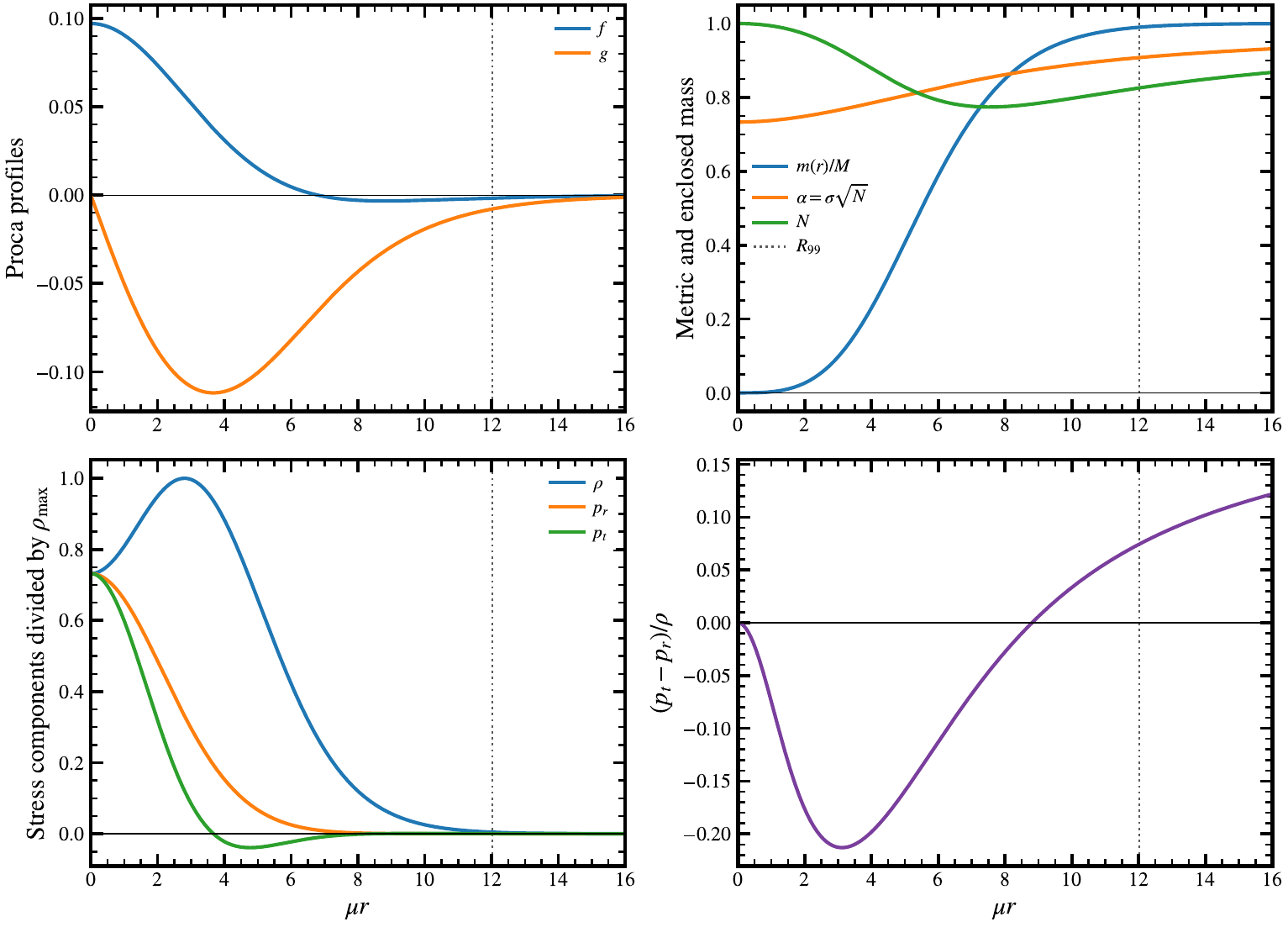}
	\caption{First-maximum solution. Top left: Proca profiles \(f\) and \(g\).
		Top right: enclosed mass, lapse \(\alpha=\sigma\sqrt N\), and \(N\).
		Bottom left: energy density and principal pressures. Bottom right: fractional
		anisotropy where \(\rho>10^{-8}\rho_{\max}\), a threshold inside the range
		over which the core extremum is invariant. The grey dotted line in every panel
		marks \(\Rnn\), enclosing \(99\%\) of the ADM mass; it is not a material
		surface.}
	\label{fig:profiles}
\end{figure*}
\section{Sequence, radii, and intrinsic stress}
\label{sec:results}

Physical masses and radii are recovered from
\begin{align*}
 M_{\rm ADM} &= 1.336\,M_\odot\,M
 \left(\frac{10^{-10}\,\mathrm{eV}}{\mu}\right),\\
 R_{\rm phys} &= 1.973\,\mathrm{km}\,R
 \left(\frac{10^{-10}\,\mathrm{eV}}{\mu}\right).
\end{align*}
For the illustrative choice \(\mu=10^{-10}\,\mathrm{eV}\), the maximum has
\(M_{\rm ADM}=1.41\,M_\odot\) and \(R_{99}=23.7\,\mathrm{km}\). This vector
mass lies above the interval excluded by the black-hole-spin analysis of
Ref.~\citep{BaryakhtarEtAl2017}, whilst a merger interpretation of an observed
transient favours a considerably lighter vector
\citep{CalderonBustilloEtAl2021}. The comparison with a neutron star depends
on the radius convention: using \(R_{99}\), the object is about twice as large
and half as compact, whereas its half-mass radius, \(R_{50}=10.9\,\mathrm{km}\),
is of neutron-star scale. Both mass and radius vary as \(\mu^{-1}\), leaving
the compactness unchanged. The conversion therefore restores physical units
but does not, by itself, constrain the particle mass.

\subsection{Profiles and force balance}
\label{sec:profiles}

Figure~\ref{fig:profiles} shows the first-maximum profiles. The temporal
component \(f\) has its single node at \(\mu r=6.79\), whereas the radial
component \(g\) is nodeless; both decay exponentially. The density is
non-monotonic, reaching \(\rho_{\max}\) at \(\mu r=2.81\) whilst its central
value is \(27\%\) smaller, a depression also reported by Lazarte and
Alcubierre \citep{LazarteAlcubierre2024}. At \(\mu r=4.70\) the density
returns to its central value, but \(\pr/\rho=0.12\) rather than unity, so no
single-valued radial barotrope \(\pr=\pr(\rho)\) can represent the solution.
Moreover, \(\pt<0\) on \(3.67<\mu r<8.41\) and \(\pr<0\) beyond
\(\mu r=10.11\). The monotonic-density and positive-pressure assumptions
often imposed on phenomenological anisotropic fluids therefore fail, even
though the dominant energy condition, Eq.~\eqref{eq:dec}, holds everywhere.

The anisotropy supplies the sharper signature. It vanishes at the centre by
regularity, remains negative throughout the core, and changes sign at
\(\mu r=8.81\), precisely where the lapse reaches
\(\alpha=\omega/\mu=0.874\), as required by Eq.~\eqref{eq:turning}. Its
largest fractional magnitude inside that surface is
\begin{equation}
\max_{0<\mu r<8.81}\frac{|\Delta|}{\rho}=0.213.
\label{eq:anisomax}
\end{equation}
This maximum occurs at \(\mu r=3.12\), where
\(\rho=0.992\rho_{\max}\). Beyond the reversal, \(\Delta/\rho\) rises
towards the exact atmospheric limit of Eq.~\eqref{eq:tail}, \(0.237\),
which exceeds the core maximum. The \(r^{-1}\) correction makes the approach
slow: the ratio is \(0.148\) at \(\mu r=20\) and \(0.181\) at
\(\mu r=30\), still below its limit at the outer boundary. This
tangentially dominated envelope contains \(9.2\%\) of the mass.

\begin{figure}[!t]
	\includegraphics[width=0.99\columnwidth]{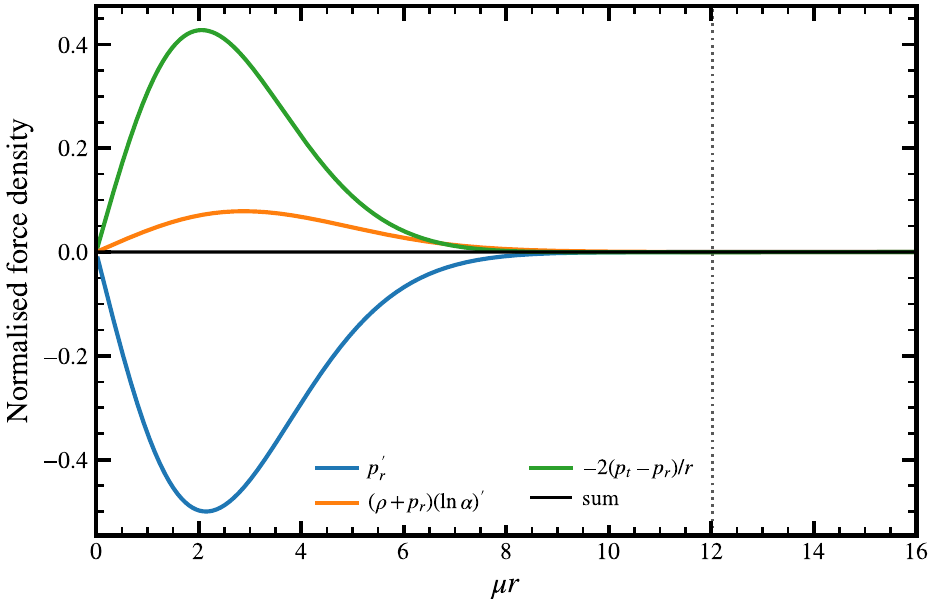}
	\caption{Terms in the covariant anisotropic balance of Eq.~\eqref{eq:atov},
		normalised by the maximum sum of their absolute values.
		The black curve is the pointwise residual. Grey dotted line: \(\Rnn\), as in
		Fig.~\ref{fig:profiles}.}
	\label{fig:forces}
\end{figure}

Figure~\ref{fig:forces} decomposes Eq.~\eqref{eq:atov} into its pressure,
gravitational and anisotropic terms out to \(\mu r=16\), with the residual
evaluated on \(0.02<\mu r<15\). The reversal has a direct mechanical
meaning: the anisotropic force density \(2\Delta/r\) points inwards in the
core and outwards in the envelope, vanishing on the lapse-defined surface.

\section{Discussion}
\label{sec:discussion}

The non-central zero provides a direct test of phenomenological closures. At
\(\mu r=8.81\), \(\Delta=0\), whilst
\(\rho=0.066\rho_{\max}\), \(\pr=0.009\rho\), and \(2m/r=0.218\);
\(\pr\) reaches zero only at \(\mu r=10.11\). The anisotropy zero is therefore
neither a pressure surface nor a consequence of vanishing matter variables.
For a closure \(\Delta=C\,\Phi(r,\rho,\pr,m/r)\), with \(C\ne0\) and
\(\Phi>0\) on the positive orthant, the four positive arguments forbid
\(\Delta=0\) at the crossing. This excludes the fixed-sign Bowers--Liang form
\citep{BowersLiang1974} and the bilinear quasi-local model of
Ref.~\citep{HorvatEtAl2011}; Ref.~\citep{HerreraSantos1997} reviews the broader
anisotropic-fluid context. The covariant prescription of
Ref.~\citep{RaposoEtAl2019} depends on the radial pressure gradient and thus
lies outside this algebraic class, but remains fixed-sign on the monotonic
equilibria studied there. The Proca identity instead contains the
lapse-dependent mass-shell factor
\(\left(\omega/\alpha\right)^2-\mu^2\): on the nodeless fundamental branch,
its zero at \(\alpha=\omega/\mu\) reverses the sign. The conclusion is
deliberately limited to the stated sign-definite closures.

The minimally coupled scalar provides a useful control case. For a neutral
complex scalar \(\Psi=e^{-i\omega t}\psi(r)\), with matter Lagrangian
\(\mathcal L_m=-\tfrac12(\partial_\lambda\Psi\,
\partial^\lambda\overline\Psi+\mu^2\Psi\overline\Psi)\), the anisotropy is
\(\Delta=-N\psi'^2\le0\)
\citep{SchunckMielke2003}, whilst in the atmosphere
\(\Delta/\rho\to-(1-\omega^2/\mu^2)\), the negative of
Eq.~\eqref{eq:tail}. Here ``neutral'' means that the conserved Noether charge
is global rather than electromagnetic, whilst ``minimal'' excludes a direct
coupling to curvature or torsion. Since \(\Delta/\rho\) is unchanged by an
overall rescaling of the Lagrangian, the comparison is convention-independent:
the scalar and Proca tails have equal fractional magnitudes, opposite signs,
and are fixed by the frequency alone.

The reversal therefore reflects the vector character of the field, rather
than merely its harmonic phase in curved spacetime. It follows from competition
between the electric-like field-strength term \(\mathcal E^2\) and the
longitudinal mass term \(\mathcal X_r\); neither \(\mathcal E^2\) nor the
global charge \(Q\) is electromagnetic. A scalar model can be made to reverse
by gauging its \(U(1)\) symmetry, which introduces a positive Maxwell
contribution to \(\Delta\), or by adding a non-minimal coupling to
teleparallel torsion, whose contribution need not have a fixed sign
\citep{HorvatEtAl2015}. The minimal complex Proca field requires neither
extension: its physical longitudinal polarisation is sufficient.

Tangential dominance itself is not unique to this system. The real-vector
stars of Ref.~\citep{Tasinato2022} attain vanishing radial pressure and high
compactness through an explicit Einstein-tensor coupling. No such non-minimal
curvature term is present here; the reversal arises within the minimal complex
Proca stress and occurs at the local mass-shell surface.

\section{Conclusions}
\label{sec:conclusions}

We have derived an exact on-shell identity for the intrinsic pressure
anisotropy of a spherical, minimally coupled complex Proca field. Its sign is
set by a local mass-shell factor, not by an independently prescribed closure.
On the nodeless fundamental branch, the radial pressure dominates the core,
the principal pressures cross once where \(\alpha=\omega/\mu\), and the
tangential pressure dominates in the atmosphere. At the first mass maximum,
the crossing occurs at \(\mu r=8.81\), the core fractional anisotropy reaches
\(0.213\), and the outer region contains \(9.2\%\) of the mass. The exact tail
ratio, \(0.237\), is the negative of the scalar limit, identifying the
longitudinal polarisation as the origin of the reversal.

Radius, density, radial pressure, and enclosed compactness remain positive at
the crossing. No sign-definite algebraic closure built from those variables
can therefore reproduce the Proca profile, although gradient-dependent and
lapse-dependent prescriptions lie outside this restricted class. Despite its
non-monotonic density and sign-changing pressures, the solution is regular,
covariantly conserved, and satisfies the dominant and strong energy
conditions. The anisotropy is thus a first-principles property of minimal
vector matter, rather than an auxiliary fluid assumption.

Global collocation and independent shooting reproduce the profiles to
\(1.8\times10^{-10}\), with local and integral residuals providing separate
checks. The first mass maximum is also the radial-stability boundary. This
well-tested stress profile now enables controlled calculations of tidal
response, mode spectra, and merger dynamics relevant to next-generation
gravitational-wave detectors and multimessenger observations. Quantifying
those signatures without a phenomenological closure is the natural next step.

\begin{acknowledgments}
The author acknowledges financial support from Funda\c{c}\~ao para a Ci\^encia
e a Tecnologia (FCT), Portugal, to the Centro de Astrof{\'i}sica e
Gravita\c{c}\~ao (CENTRA) through grants UID/PRR/00099/2025
(\doi{10.54499/UID/PRR/00099/2025}) and UID/00099/2025
(\doi{10.54499/UID/00099/2025}).
\end{acknowledgments}

\bibliography{Art_ILopes}

\appendix

\section{Primary global collocation eigenvalue formulation}
\label{app:collocation}

The collocation calculation uses the asymptotic-time gauge,
\(\sigma(r_{\max})=1\), so \(\omega\) is the physical frequency of the
finite-domain problem and \(\kappa=\sqrt{1-\omega^2}\). The continuation
parameter \(a_0\), introduced in Sec.~\ref{sec:numerics}, is gauge invariant.
At fixed \(a_0\), the four first-order fields
\(\bm y=(m,\sigma,f,g)\) and the eigenvalue \(\omega\) satisfy
\begin{align}
m(0)&=0,\qquad g(0)=0,
\nonumber\\
f(0)-a_0\sigma(0)&=0,\qquad \sigma(r_{\max})=1,
\nonumber\\
f'(r_{\max})+
\left(\kappa+\frac{1}{r_{\max}}\right)f(r_{\max})&=0.
\label{eq:collocation_bc}
\end{align}
The Robin condition on \(f\) removes the growing asymptotic mode and closes the
nonlinear eigenvalue problem. Prescribing both \(f(0)=a_0\) and
\(\sigma(r_{\max})=1\) would instead combine the central-time and
asymptotic-time normalisations, thereby changing the meaning of \(a_0\).

Because Eq.~\eqref{eq:gprime} is singular at \(r=0\), the collocation domain
begins at \(r_0>0\). Regularity is imposed through the central expansion,
rather than through the limiting values of Eq.~\eqref{eq:collocation_bc}.
In the dimensionless variables of Sec.~\ref{sec:numerics}, define
\(\sigma_c=\sigma(0)\), whence \(f_c=a_0\sigma_c\). The inner data are
\begin{align}
m(r_0)&=\frac{a_0^2}{6}r_0^3+\mathcal O(r_0^5),
\nonumber\\
\sigma(r_0)&=\sigma_c\left(1+\frac{a_0^2}{2}r_0^2\right)
+\mathcal O(r_0^4),
\nonumber\displaybreak[4]\\
f(r_0)&=a_0\sigma_c\left[
1+\frac16\left(1-\frac{\omega^2}{\sigma_c^2}\right)r_0^2
\right]+\mathcal O(r_0^4),
\nonumber\\
g(r_0)&=-\frac{a_0\omega}{3\sigma_c}r_0+\mathcal O(r_0^3).
\label{eq:collocation_centre}
\end{align}
The scalar pair \((\omega,\sigma_c)\) completes the unknown set. The four
central-series residuals in Eq.~\eqref{eq:collocation_centre} and the two outer
residuals in Eq.~\eqref{eq:collocation_bc} provide the required six
constraints.

The spectral structure is exposed by holding the geometry fixed. With
\(H=(f'-\omega g)/\sigma\) and \(u=r^2H\), the two Proca equations combine, in
the same dimensionless units, into
\begin{equation}
-\frac{\dd}{\dd r}\left(\frac{\sigma N}{r^2}\frac{\dd u}{\dd r}\right)
+\frac{\sigma}{r^2}u
=\omega^2\frac{u}{\sigma N r^2},
\label{eq:collocation_sl}
\end{equation}
a singular Sturm--Liouville problem for \(\omega^2\). In the full system,
however, \(N\) and \(\sigma\) are generated by the same eigenfunction; the
problem is therefore a self-consistent nonlinear eigenvalue problem, rather
than a linear fixed-background spectrum.

The resulting algebraic system is solved by a damped Newton method and
continued in \(a_0\). A shooting profile initialises only the first
collocation solve; no shooting value enters the converged residuals, boundary
conditions, or tabulated observables.

\newpage

\section{Independent shooting construction}
\label{app:shooting}

The shooting calculation provides an independent numerical check of the
collocation scheme in Appendix~\ref{app:collocation}. It solves the same
first-order radial system, but as an outward initial-value problem:
Eq.~\eqref{eq:proca2} eliminates \(f'\), and the remaining equations are
integrated from the regular expansion at \(r_0=10^{-6}\) to \(r_{\max}=35\).
Equation~\eqref{eq:proca1} is deliberately withheld from the integration and
is subsequently evaluated as a residual.

The distinction between the two methods is straightforward. Collocation
determines the complete profile and eigenfrequency simultaneously in the
asymptotic-time gauge. Shooting instead adopts the central-time gauge
\(\sigma_c=1\), integrates outwards, and adjusts the trial frequency until the
same outer decay condition is met. The resulting solution is then rescaled to
the asymptotic-time gauge; this changes \(\sigma\), \(f\), and \(\omega\), but
leaves the enclosed mass and charge invariant. The integration uses an
adaptive eighth-order Dormand--Prince method, with relative and absolute
tolerances of \(2\times10^{-10}\) and \(10^{-12}\). Residual derivatives are
reconstructed independently from splines of the stored profiles.

At matched amplitudes, representative dilute, maximum-mass, and dense
solutions agree with the collocation results within \(2.7\times10^{-10}\).
The two methods locate the mass maximum independently, at amplitudes differing
by \(9.4\times10^{-8}\). At their respective maxima, the differences in \(M\)
and \(Q\) are of order \(10^{-11}\), whilst those in \(\omega\), the central
lapse, and \(R_{99}\) are \(3.6\times10^{-8}\), \(1.0\times10^{-7}\), and
\(3.1\times10^{-6}\), respectively. The coincident entries in
Table~\ref{tab:benchmark} therefore reflect rounding, not identical numerical
solutions.

At a common amplitude near the maximum, the two profiles differ by
\(1.8\times10^{-10}\). The off-mesh collocation and local field residuals lie
below \(10^{-7}\), the hydrostatic residual below \(10^{-6}\), and the virial
residual is of order \(10^{-10}\) for both methods. Increasing \(r_{\max}\)
from 35 to 40 changes \(M\) and \(Q\) fractionally by less than
\(6\times10^{-11}\). Thus the agreement between
Appendices~\ref{app:collocation} and~\ref{app:shooting} tests two independent
constructions of the same boundary-value problem. Since both use the same
field equations and approximate outer condition, the separate
\(r_{\max}\)-variation test is needed to verify that the finite boundary does
not control the reported observables.

\end{document}